\documentclass[runningheads]{svjour2}
\smartqed  
\usepackage{graphicx, epsfig, amssymb} 
\usepackage{amsmath, amsfonts}
\usepackage{bm} 
\usepackage[linktocpage]{hyperref}
\usepackage[caption=false]{subfig}
\usepackage[usenames]{color}
\newcommand{\be}{\begin{equation}}
\newcommand{\ee}{\end{equation}}
\newcommand{\beq}{\begin{eqnarray}}
\newcommand{\eeq}{\end{eqnarray}}

\newcommand{\bea}{\begin{eqnarray}}
\newcommand{\eea}{\end{eqnarray}}
\newcommand{\ben}{\begin{enumerate}}
\newcommand{\een}{\end{enumerate}}
\newcommand{\bi}{\begin{itemize}}
\newcommand{\ei}{\end{itemize}}

\journalname{General Relativity and Gravitation}
\begin{document}

\title{Black hole bombs and explosions}
\subtitle{from astrophysics to particle physics}


\author{Vitor Cardoso        
}


\institute{Vitor Cardoso \at
              CENTRA, Departamento de F\'{\i}sica, Instituto Superior T\'ecnico,\\
              Universidade T\'ecnica de Lisboa - UTL, Av.~Rovisco Pais 1, 1049 Lisboa, Portugal \\
              \email{vitor.cardoso@ist.utl.pt}             \\
              \emph{Present address:} Perimeter Institute for Theoretical Physics,
              Waterloo, Ontario N2J 2W9, Canada
}



\maketitle

\begin{abstract}
Black holes are the elementary particles of gravity, the final state of sufficiently massive stars and of
energetic collisions. With a forty-year long history, black hole physics is a fully-blossomed field which 
promises to embrace several branches of theoretical physics. Here I review the main developments in highly 
dynamical black holes with an emphasis on high energy black hole collisions
and probes of particle physics via superradiance. 
This write-up, rather than being a collection of well known results, is intended to highlight open issues and 
the most intriguing results.
\end{abstract}
%
%
%

\section{Introduction}
The Kerr-Newman family of black holes in stationary four-dimensional, asymptotically
flat spacetimes exhausts all possible electro-vacuum solutions in General Relativity.
Theoretical aspects of black hole physics were fully developed decades ago,
when the Kerr-Newman family was discovered and characterized. Most of the tools to understand black hole physics
have been in place since that period and most processes involving black holes were controlled at the order-of-magnitude level for at least two decades.
Perhaps due to these reasons, ``black hole physics'' conjures up images of horizons and time-warps and old-fashioned, 
frozen-in-time topics to lay audiences.
However, in the last years the activity in the field is bubbling up and the interest in these issues has widened, 
driven by several unrelated events at the observational/instrumental, technical and conceptual levels such as 

\noindent {\it \small (i)} our capacity to observationally scrutinize the region close to the horizon within a few Schwarzschild radii, with radio and deep infrared interferometry, of which we have seen but the first steps \cite{Doeleman:2008qh,Fish:2010wu,Eisenhauer:2008tg,Antoniadis:2013pzd};

\noindent {\it \small (ii)} the ability to measure black hole spins more accurately than ever before using X-ray spectroscopy \cite{Reynolds:2013qqa,Risaliti:2013cga}
or ``continuum-fitting'' methods \cite{McClintock:2011zq}. Measurement of black hole mass {\it and} spin is one necessary step
in testing General Relativity \cite{Berti:2009kk};

\noindent {\it \small (iii)} huge technological progress in gravitational-wave observatories, some of which had been gathering data at design sensitivities for several years and are now being upgraded to sensitivities one order of magnitude higher (black hole binaries are thought to be among the first objects to ever be detected in the gravitational-wave spectrum);

\noindent {\it \small (iv)} our ability to numerically evolve black hole binaries at the full nonlinear level \cite{Pretorius:2005gq,Baker:2005vv,Campanelli:2005dd} and its immediate importance for gravitational-wave searches and high-energy physics \cite{Cardoso:2012qm};

\noindent {\it \small (v)} improvement of perturbative schemes, either by an understanding of regularization schemes to
handle the self-force \cite{Poisson:2011nh,Barack:2009ux}, or by faster and more powerful methods to deal with the full ladder of perturbation
equations \cite{Berti:2009kk,Pani:2011xj,Pani:2012bp,Pani:2013pma} (describing wave phenomena, extreme-mass ratio inspirals, etc);
      
In parallel with these technical developments, other revolutions were taking place at the conceptual level, in particular:

\noindent {\it \small (vi)} the gauge/gravity duality relating field theories to gravitational physics in anti-de Sitter spacetimes via holography \cite{Maldacena:1997re,Klebanov:2000me}. Black holes play the very important role of thermal states in this duality. The gauge/gravity duality opens up a whole new framework to understand traditionally very complicated phenomena through black hole physics \cite{Cardoso:2012qm};

\noindent {\it \small (vii)} extensions of the Standard Model to encompass fundamental ultra-light scalar fields, either minimally coupled
or coupled generically to curvature terms. These theories include, for instance, generalized scalar-tensor theories \cite{Fujii:2003pa} and the ``axiverse scenario'' 
\cite{Arvanitaki:2009fg,Arvanitaki:2010sy,Kodama:2011zc}. Ultralight scalars lead to interesting new phenomenology with possible smoking gun effects in black hole physics \cite{Cardoso:2011xi,Pani:2012vp,Yoshino:2012kn}, and are a healthy and ``natural'' extension of GR;
  
\noindent {\it \small (viii)} The formulation of TeV-scale gravity scenarios, either with warped or flat extra-dimensions \cite{ArkaniHamed:1998rs,Randall:1999ee}, most of which predict
black hole formation from particle collisions at scales well below the ``traditional'' Planck scale (see for instance Chapter 4 in Ref.~\cite{Cardoso:2012qm}).

In brief, a second ``Golden Era'' in gravitational and black hole physics has begun and it promises to shed light over a larger portion of the scientific building. The various reviews that have come out in the last couple of years on the subject are natural outcomes of the tremendous excitement in the air \cite{Cardoso:2012qm,Sperhake:2011xk,Sperhake:2013qa,Berti:2013uda,Rezzolla:2013gwa}.

\section{Superradiance and black hole physics}
The defining property of black holes is the event horizon, a one-way viscous membrane from which nothing escapes (at the classical level). Black holes are perfect absorbers.
The addition of rotation introduces another important player, a negative-energy region called the ergoregion \cite{Cardoso:2012zn}. The ergoregion is delimited by a static-limit, infinite redshift ergo-surface outside the horizon, where objects are forced to co-rotate with the black hole: rotating black holes are also perfect ``draggers'' \cite{Cardoso:2012zn}. The existence of negative-energy states {\it and} an horizon allows for an interesting effect in black hole physics, the Penrose process \cite{Penrose:1971uk,Penrose:1969pc,wald:penrose} and its wave counterpart, superradiant effects \cite{zeldovich,Bekenstein:1973mi}: 
in a scattering experiment of a wave with frequency $\omega<m\Omega$ (with $m$ an azimuthal wave quantum number and $\Omega$ the angular velocity of the horizon) the scattered wave will be amplified, the excess energy being withdrawn from the object's rotational energy \cite{zeldovich,Bekenstein:1973mi,Teukolsky:1974yv}. For reviews on superradiance in the context of the Klein-paradox I refer the reader to
Manogue's review \cite{Manogue} and for a comprehensible work on superradiance in physics to Bekenstein and Schiffer
\cite{Bekenstein:1998nt}.

\subsection{Black hole fission?}
\begin{figure}[ht]
\begin{center}
\epsfig{file=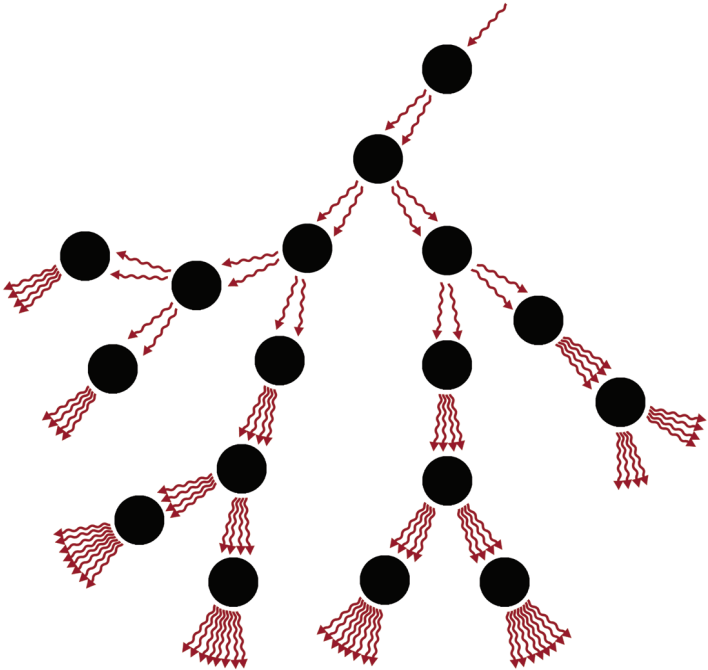,width=10cm,angle=0}
\caption{Scheme of the hypothetical chain reaction in a cluster of rotating black holes. The incident
arrow denotes an incident wave on the rotating black hole, which is then amplified and exits with larger amplitude, before interacting with other black holes.
The super-radiantly scattered wave interacts with other black holes, in an exponential cascade.
\label{fig:chainreaction}}
\end{center}
\end{figure}

Superradiance can be used to tap energy from black holes in several different ways. The most naive method for energy extraction
would consist on scattering a bosonic wave off a Kerr black hole. Unfortunately, typical amplification factors are small \cite{Teukolsky:1974yv},
and many scatters are required. A more ingenious approach would be to build a kind of fission apparatus, depicted in Fig.~\ref{fig:chainreaction}.
We take a cluster of rotating black holes, and send in a low-frequency photon. If the cluster is appropriately built, it would seem {\it possible in principle} that the photon is successively amplified as it scatters off, leading to an exponential cascade. This kind of process is identical to the way fission bombs work, where neutrons play the role of our wave.

It was pointed out by Press and Teukolsky~\cite{Teukolsky:1974yv} that such a process could not occur for Kerr black holes, as the entire cluster would have to be contained in its own Schwarzschild radius. Let us see how this works in a generic $D$-dimensional setting. We take a cluster of $N$ rotating black holes of size $L$, and total mass $NM_{BH}$, where $M_{BH}$ is the mass of each individual black hole. Assuming all the conditions are ideal, the process can only work if the mean free path $\ell$ of a photon (or any other boson field) is smaller than the size of the cluster,
\be
\ell<L\,.
\ee
Now, the mean free path is $\ell=\frac{1}{n\sigma}$, where $n$ is the black hole number density in the cluster and $\sigma$ is an absorption cross section. The absorption cross-section is {\it at best} negative if a plane wave is amplified upon incidence on a rotating black hole.
Even in such case, it is of order the black hole area. 
These two properties are very important. That the cross-section scales with the area can be seen on purely dimensional arguments
and it holds true for all black hole spacetimes we know of. A negative total cross-section is necessary to guarantee that whatever way the boson is scattered it will {\it on the average} be superradiantly amplified. In other words, we require that a plane wave is subjected to superradiance.
\footnote{It is sometimes not appreciated that a, say, $l=m=1$ mode is a sum of modes with respect to some other coordinate frame, where the following (black hole) scatterer is sitting.}
As far as I'm aware, none of the known black hole geometries have a negative cross-section. This would probably lead to an instability of
the single black hole itself, although {\it that} is of no concern for us here. Attempts at building analog fluid geometries with large amplification factors were successful, but negative cross-sections were never seen \cite{Marques:2011}.
To summarize,
\be
\sigma \sim V_{D-2}r_+^{D-2}\,,
\ee
where $V_{D-2}=\pi^{D/2-1}/\Gamma[D/2]$ is the volume of a unit $(D-3)$ sphere. Thus, up to factors of order unity the condition for fission would amount to $L^{D-2}/(N r_+^{D-2})<1$ or equivalently
\be
\frac{NM_{BH}}{L^{D-3}}>\frac{L}{r_+}\,.
\ee
This last condition is stating that the cluster lies within its own Schwarzschild radius, making the fission process impossible even in the most idealized scenario.
\subsection{Black hole bombs}
%
\begin{figure}[ht]
\begin{center}
\epsfig{file=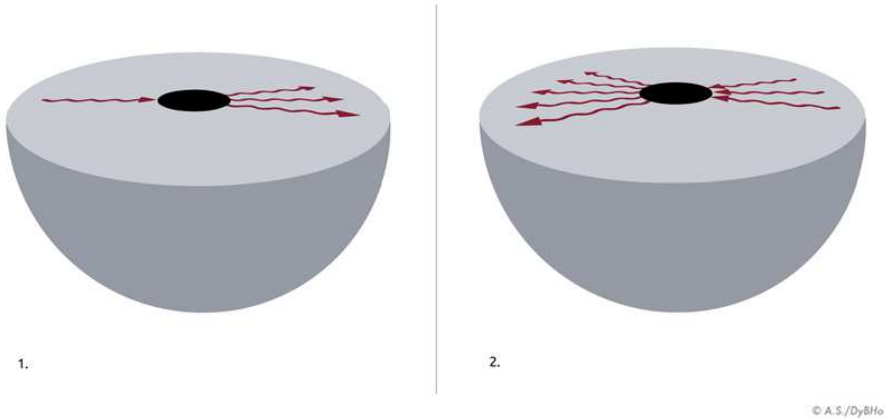,width=12cm,angle=0}
\caption{Rotating black hole surrounded by perfectly reflecting cavity. Low-frequency radiation is successively reflected at the mirror
and amplified close to the black hole, in an exponential cascade.
\label{fig:bhbomb}}
\end{center}
\end{figure}
Another simple way to tap the hole's rotation energy via superradiance is to enclose the rotating black hole inside a perfectly reflecting cavity, as in Fig.~\ref{fig:bhbomb}. Any initial perturbation will get successively amplified near the black hole and reflected back at the mirror, thus creating an instability, which was termed the ``black hole bomb'' by Press and Teukolsky \cite{Press:1972zz}, and explored in detail
in Refs.~\cite{Cardoso:2004nk,Strafuss:2004qc,Dolan:2012yt,Hod:2009cp,Rosa:2009ei}. Such a device could be used to extract the energy piled up in the field, for instance in the radio-band \cite{Press:1972zz}. Extensions to higher number of spatial dimensions was performed in Ref.~\cite{Lee:2011gt}.

\begin{figure}[ht]
\begin{center}
\includegraphics[width=0.48\textwidth]{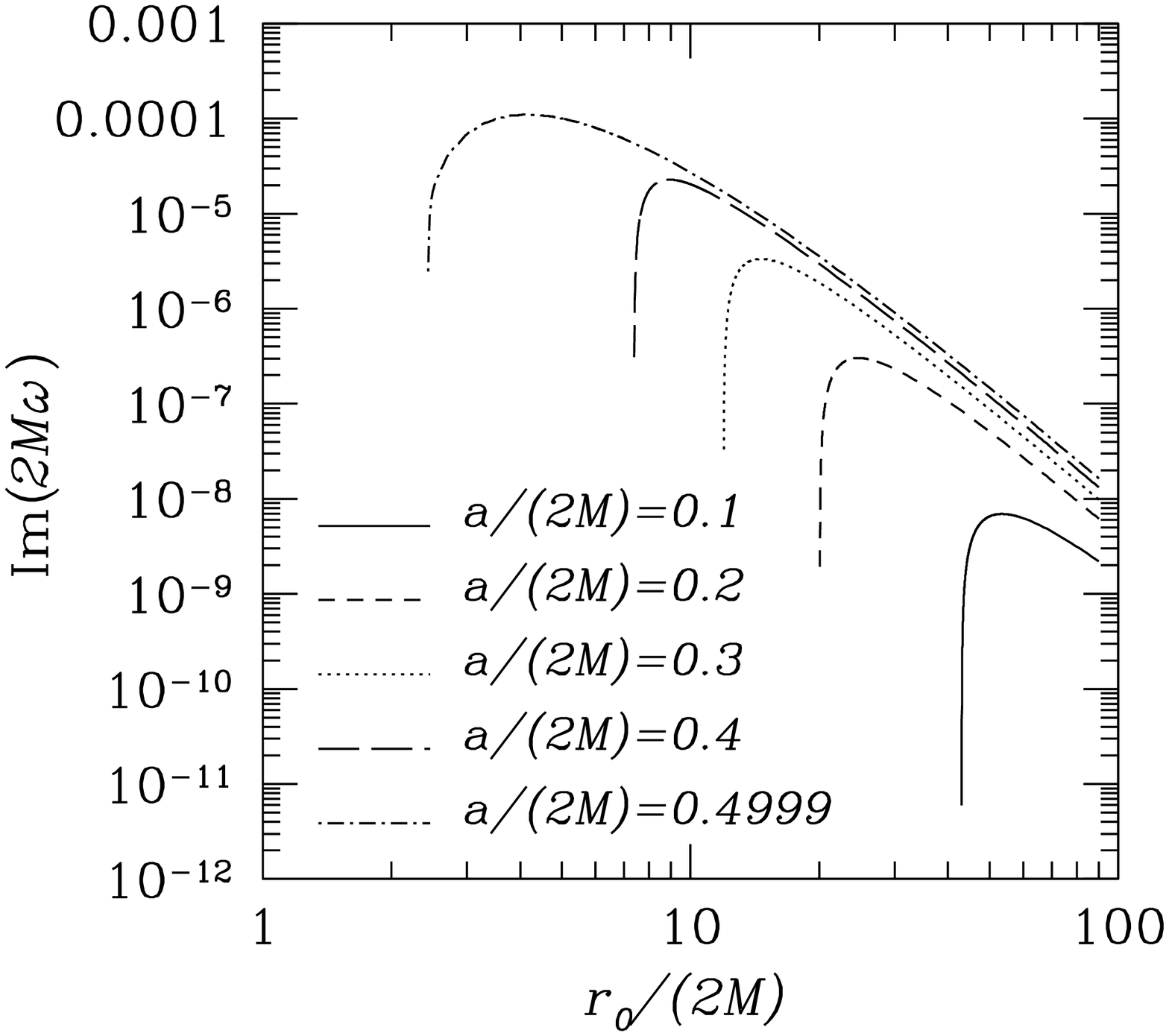}
\includegraphics[width=0.48\textwidth]{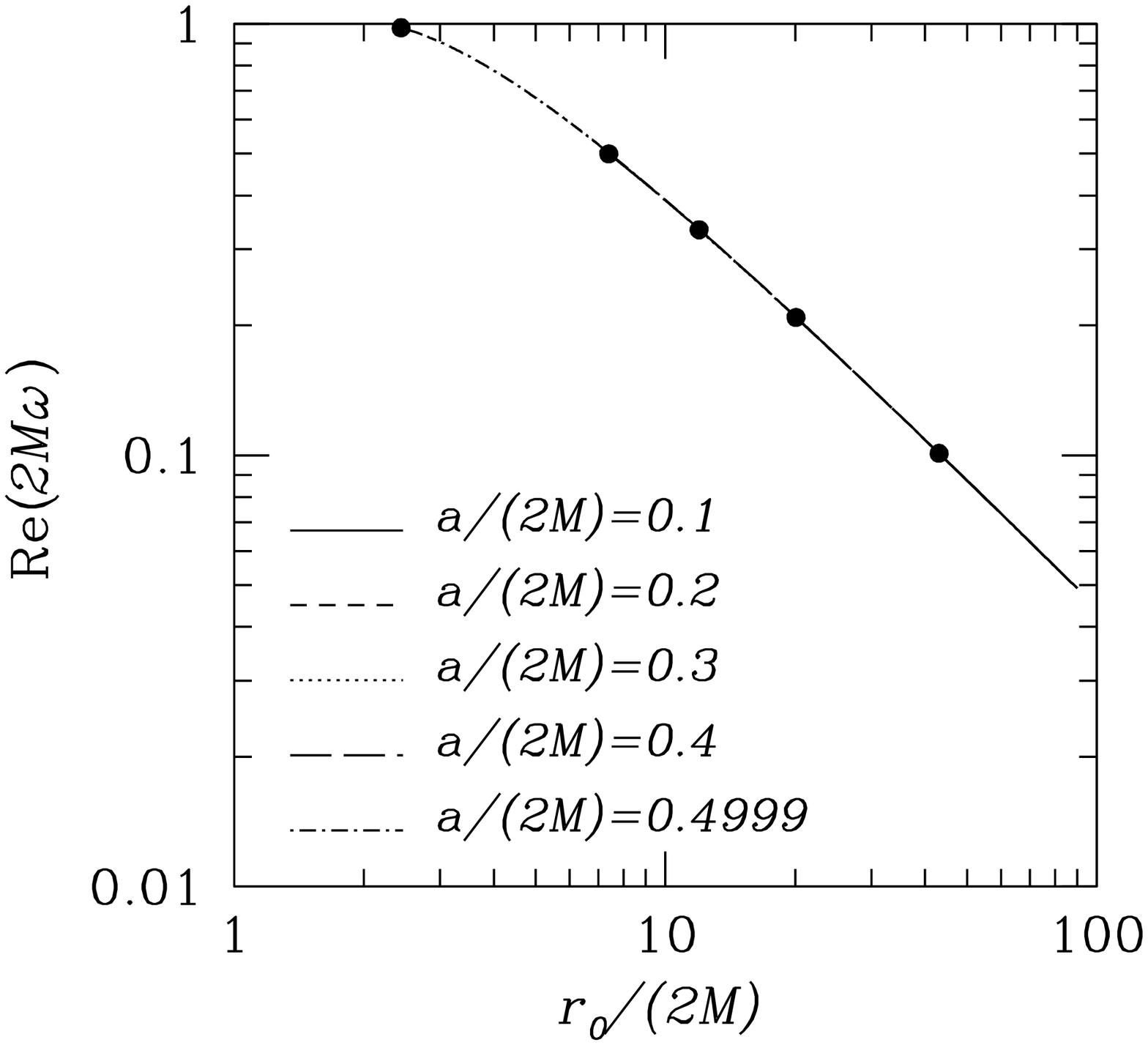}
\caption{Rotating black hole surrounded by perfectly reflecting cavity. Low-frequency radiation is successively reflected at the mirror
and amplified close to the black hole, in an exponential cascade. Taken from Ref.~\cite{Cardoso:2004nk}.
\label{fig:bhbombnumerics}}
\end{center}
\end{figure}

The scalar-field black hole bomb was worked out in Ref.~\cite{Cardoso:2004nk} and is summarized in Fig.~\ref{fig:bhbombnumerics}; here we show the real and imaginary component of the resonant frequencies in the system,
for a scalar with time dependence $\Phi\sim e^{-i\omega t}$. 
These results are very instructive: the ringing frequencies (ie., the real part of the modes) show a very weak dependence on the rotation rate of the black hole. In fact, the ringing frequencies are dictated only by the mirror radius. For small mirror radius the superradiant instability is quenched, as the natural frequencies of the system (which are inversely proportional to the system's size) are very large and do not satisfy the superradiant condition $\omega>\Omega$. Finally, it should be noted that the typical growth scale (or decay, when the modes are not superradiant) of perturbations in a cavity of radius $r_0$ scale as \cite{Cardoso:2004nk}
\be
\tau \sim r_0^{2(l+1)}\,, \label{tau_dependence}
\ee
for scalar-field perturbations; such modes are therefore extremely long-lived.

The extremely long timescales necessary for superradiant instabilities to develop are a nuisance from
a numerical perspective, as very long and accurate simulations are necessary. It was very recently shown that superradiant bombs can become very efficient, if instead of rotation one considers charged black holes and charged scalar fields \cite{Herdeiro:2013pia,Zhang:2013haa}. Although astrophysical black holes are expected to be neutral, or very nearly so,
this would perhaps constitute an interesting toy model for which nonlinear evolutions are feasible.

\subsection{Black hole bombs in astrophysics: accretion disks and torus}
Scalar fields are a nice prototype, but there is only so much they can tell us about astrophysical processes involving electromagnetic (EM) 
interactions. However, black holes are typically surrounded by plasmas, which reflect low-frequency electromagnetic (EM) waves. 
Thus, there are two questions that should be addressed. The first concerns the process itself: are black hole bombs still operative in the EM spectrum? 
The second question concerns realistic walls: it is likely that the matter surrounding the black hole comes under the form of thin or thick accretion disks
and not as spherical mirrors. Is this enough to completely wash out this black hole bomb effect?
Finally, are these putative instabilities easily quenched by losses in the surrounding ``mirror''?

These are questions which have been dealt with in the past \cite{Putten,Aguirre:1999zn}, but which need
serious work on them to be fully understood. Electromagnetic black hole bombs are 
argued to be conceptually efficient because the maximum superradiance amplification factor for EM waves is one order of 
magnitude larger than for scalar waves \cite{Teukolsky:1974yv}.
In the bounce-and-amplify picture of the black hole bomb, the instability timescale should accordingly be one order of magnitude shorter for EM
waves. However, this picture is very particle-like, whereas the unstable modes are very low frequency, so that a wave analysis
is imperative. In fact, it is hard to reconcile a particle-like picture with the dependence in Eq.~(\ref{tau_dependence}) (the timescale should be proportional to a light-travel time and hence to $r_0$). A rigorous analysis of the EM black hole bomb is necessary.

The second question concerns the non-sphericity of the surrounding matter. If the instability develops {\it within}
an accretion disk or torus, it has been argued the instability is weakened by geometrical constraints \cite{Aguirre:1999zn}:
it has been shown that higher-angular-eigenvalue modes have longer instability timescales when the mirror
is spherical \cite{Cardoso:2004nk}. Confining the field further along some angular direction effectively
means the lower-angular-eigenvalue modes are forbidden. 

The geometrical constraint imposed by torus or accretion disks does not kill the instability, but it has been argued that absorption effects do
\cite{Aguirre:1999zn}. The argument runs as follows: in each scatter off the central black hole, the wave gets amplified by roughly $\sim 1\%$.
After reflection at the cavity wall, there is a net gain only if the wall is $99\%-$ dissipation free or higher \cite{Aguirre:1999zn}, which 
is hard to justify. However, superradiance thrives in dissipation: the ``wall'' is most likely to be made of material rotating at high velocities
around the central black hole. A small absorption will cause further {\it amplification} at the wall!
In fact, in a flat-space calculation, Bekenstein and Schiffer \cite{Bekenstein:1998nt} consider a rotating cylinder with angular velocity $\Omega$ and radius $R$, made of material with spatially uniform permittivity $\epsilon(\omega)$, permeability $\mu(\omega)$
and conductivity $\sigma$. They find a maximum amplification factor of 
\be
\frac{{\rm Flux}_{\rm out}}{{\rm Flux}_{\rm in}}\sim 1.185 \left(\omega R/c\right)^2 \,\qquad {\rm at}\qquad \Omega=\omega+0.503c^2/\sigma R^2\,.
\ee
For realistic accretion disks or torus around black holes, $\omega R/c$ can be of order unity, making superradiance
at the cavity wall dominant over that at the black hole.

It is even possible that, under the right circumstances, the presence of matter stimulates black hole bombs: Pani and Loeb have recently shown that
the plasma surrounding primordial black holes acts as an effective mass for electromagnetic waves, which get confined and grow exponentially
in the vicinity of these holes \cite{Pani:2013hpa}. Subsequent evolution of the system may leave an observabe imprint on the cosmic microwave background.
\subsection{Black hole bombs in anti-de Sitter space}
Anti-de Sitter spacetimes are a very natural realization of the black hole bomb instability, as their timelike boundary is perfectly suited to play the role of the reflecting cavity. Indeed, it was first shown in Ref.~\cite{Cardoso:2004hs} that small rotating anti-de Sitter black holes are unstable against a superradiant mechanism. This study was generalized to other geometries and perturbation sectors \cite{Cardoso:2006wa,Kodama:2007sf,Uchikata:2009zz,Murata:2008xr,Kodama:2009rq}. A full analysis of gravitational perturbations in four-dimensions is underway \cite{Cardoso:2013}. Large black holes in anti-de Sitter are {\it not} superradiantly unstable \cite{Hawking:1999dp}.

In asymptotically flat spacetimes, the endpoint of the superradiant instability is necessarily a Kerr black hole with lower angular momentum.
Physically this is because the instability is non-axisymmetric and any such perturbation radiates gravitational waves which carry angular momentum.
In anti-de Sitter however, the boundary can prevent leakage of energy and angular momentum, and superradiance onset can in principle
also signal bifurcation to other black hole families, possibly hairy ones.
Such possible end-states were discussed in Refs.~\cite{Cardoso:2006wa,Basu:2010uz,Dias:2011tj,Gentle:2011kv}.
Some examples of rotating solutions with scalar hair were explicitly constructed in Refs.~\cite{Dias:2011at,Stotyn:2011ns}.

\subsection{Massive fields, soft bombs and particle physics}
%
\begin{figure}
\begin{center}
\begin{tabular}{cc}
\includegraphics[width=0.5\textwidth]{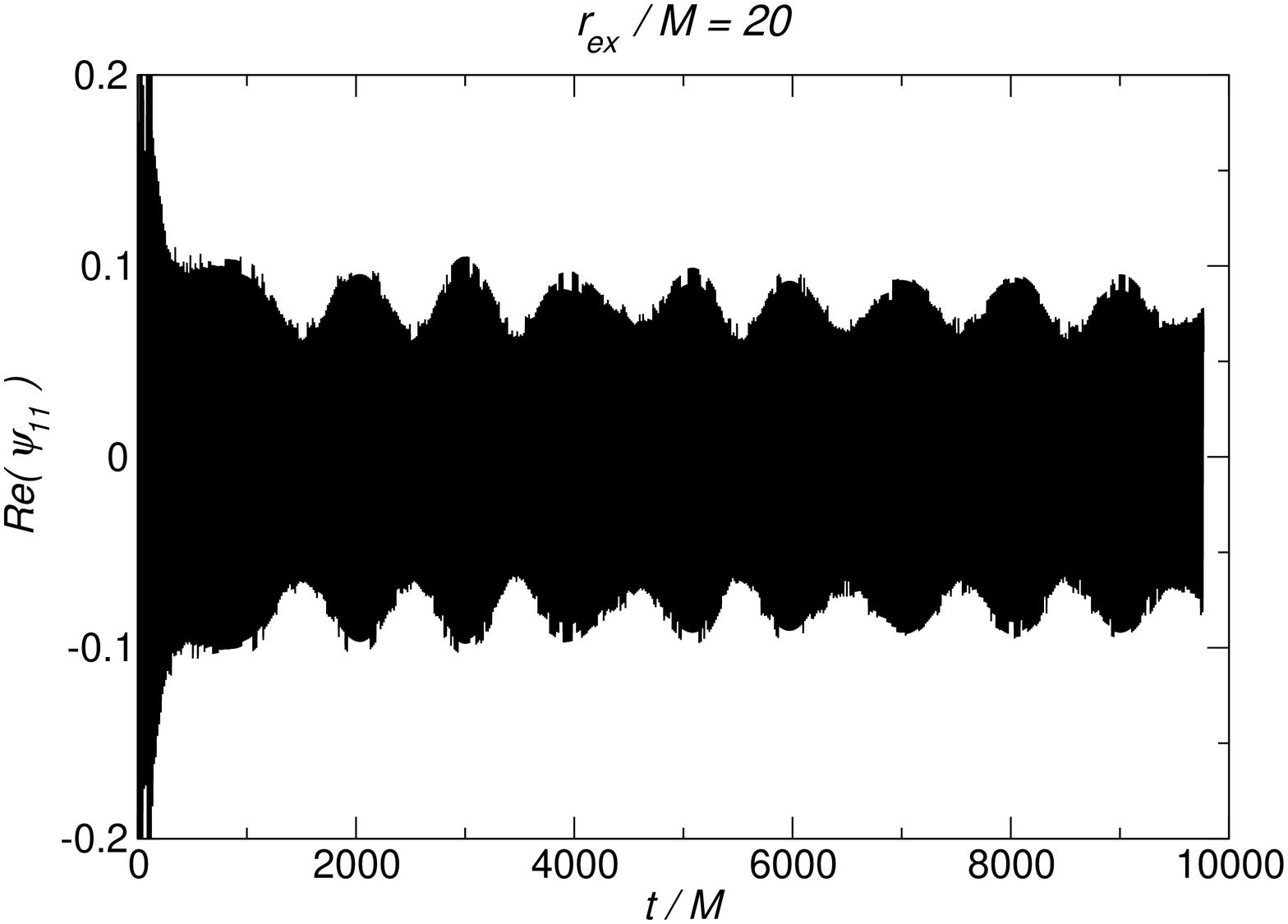} &
\includegraphics[width=0.5\textwidth]{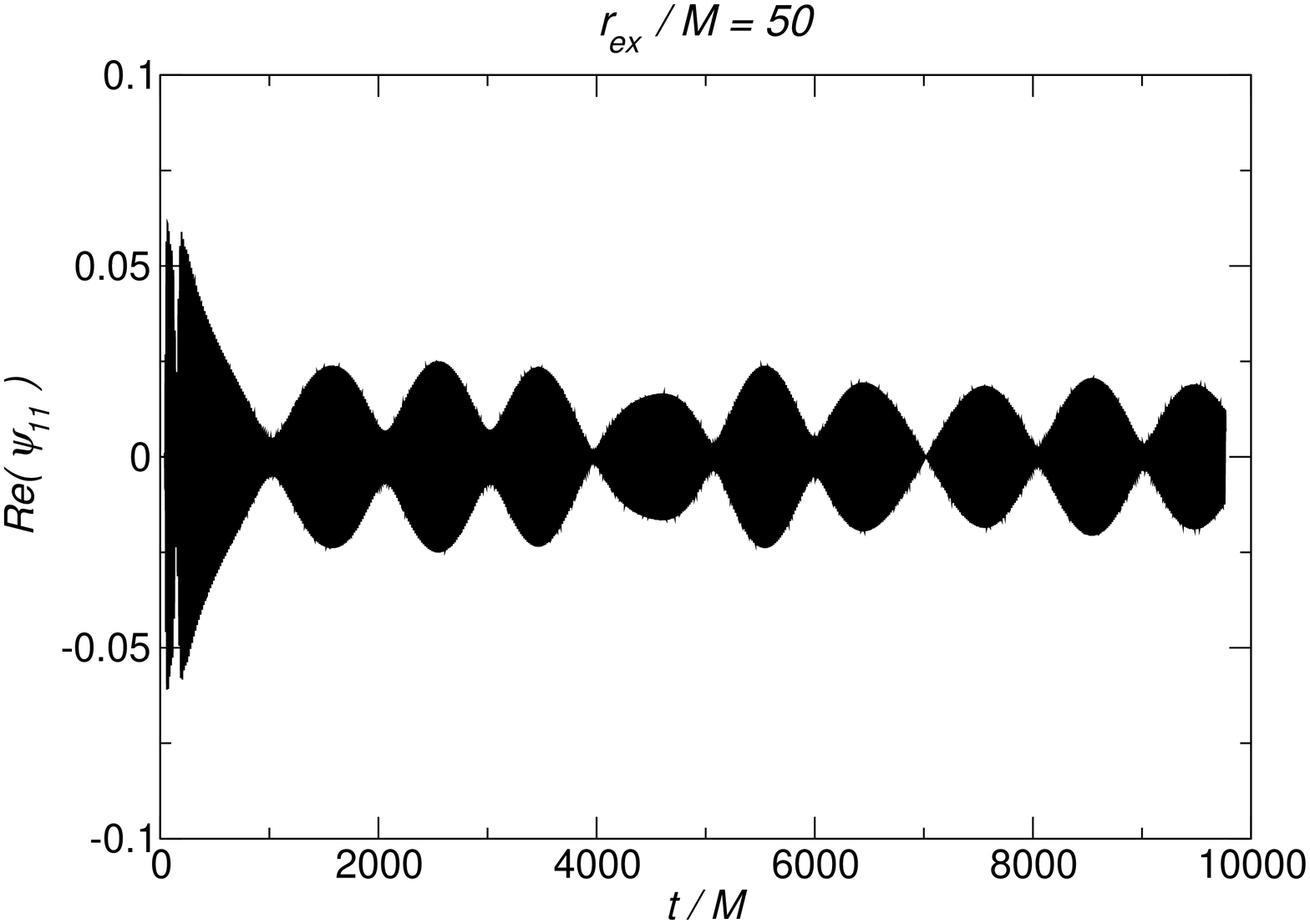} \\
\end{tabular}
\end{center}
\caption{\label{fig:waveformsSSm042_11}
The dipole amplitude at selected extraction
radii $r_{\rm ex}$ for a massive scalar field with $M\mu_S = 0.42$ in
a Kerr background with $a=0.99M$. The initial pulse is a gaussian $\Psi_{11}=e^{-(r-12)^2/4}$
and has angular dependence given by $Y_{1-1}-Y_{11}$ with $Y_{lm}$ being the scalar spherical harmonic. 
The black hole has mass $M=1$. Taken from Ref.~\cite{Witek:2012tr}.
}
\end{figure}
Massive fields are another realization of a natural reflecting wall, opening up the exciting possibility to test particle physics via black hole physics 
\cite{Arvanitaki:2010sy,Pani:2012bp,Pani:2012vp}. The idea is very simple and consists in thinking about
these massive states orbiting the black hole in bound {\it stable} orbits while extracting energy from the ergoregion in an exponentially growing way \cite{Damour:1976kh}. Because there {\it are} stable orbits of timelike particles in Kerr, this analogy would imply that the Kerr spacetime
is unstable against massive field perturbations. The instability has been checked analytically and numerically for scalar
\cite{Damour:1976kh,Zouros:1979iw,Detweiler:1980uk,Furuhashi:2004jk,Cardoso:2005vk,Dolan:2007mj,Hod:2012px}, vector \cite{Rosa:2011my,Pani:2012bp,Pani:2012vp,Witek:2012tr}
and tensor fields \cite{Brito:2013wya}\footnote{Note that Kaluza-Klein compactifications in extra-dimensional frameworks are equivalent to
effective masses \cite{Cardoso:2004zz,Rosa:2012uz}.}, and the first rigorous construction of a superradiant
instability appeared very recently \cite{Shlapentokh-Rothman:2013ysa}.

In Fig.~\ref{fig:waveformsSSm042_11} I show how the instability could develop for generic initial data \cite{Witek:2012tr}.
The figure refers to a scalar field, but the main qualitative features should be universal. The initial data is a gaussian pulse
evolved on a nearly-extreme Kerr black hole geometry (see Ref.~\cite{Witek:2012tr} for details). The instability is not apparent. 
In fact, longer-lasting evolutions would be required for such instability to become visible.
Its interesting feature is the presence of beating patterns due to the hydrogenic-like spectrum of the massive states.

\begin{figure}[htb]
\begin{center}
\epsfig{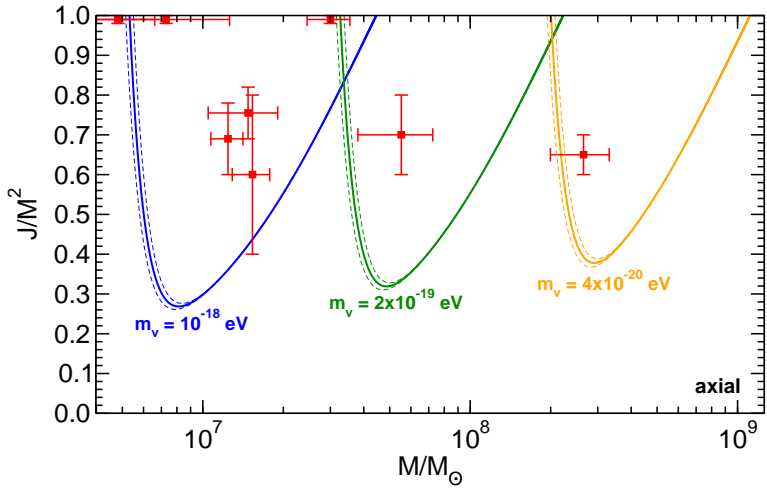}
\caption{Contour plots in the black hole Regge plane~\cite{Arvanitaki:2010sy}
  corresponding to an instability timescale shorter than $\tau_{\rm
    Hubble}$ (continuous lines) or $\tau_{\rm Salpeter}$ (dashed
  lines) for different values of the vector field mass
  $m_v={{\mu}}\hbar$. The experimental points (with error bars) refer to
  the supermassive black holes listed in Table~2 of~\cite{Brenneman:2011wz};
  the rightmost point corresponds to the supermassive black hole in Fairall 9
  \cite{Schmoll:2009gq}. Supermassive black holes lying above each of these
  curves would be unstable on an observable timescale, and therefore
  they exclude the corresponding range of Proca field
  masses. Taken from Refs.~\cite{Pani:2012bp}.\label{fig:bound2}}
\end{center}
\end{figure}

The superradiant instability of massive fields has become more relevant in the context of the ``axiverse`'' scenario, where boson fields acquire a (very) small mass, and which are not ruled out by experiments~\cite{Arvanitaki:2009fg,Arvanitaki:2010sy}. The {\it observation}
of rotating black holes allows (in principle, see below) to impose interesting constraints on the mass of long-range bosonic fields.
The idea is as follows: the instability is active on a timescale $M\tau \sim (M\mu)^{-\alpha}$, where $\alpha=9$ for scalar fields for instance 
\cite{Pani:2012bp,Pani:2012vp,Witek:2012tr}. The only stationary vacuum solution of Einstein equations is the Kerr family; accordingly, the end-state of such mechanism must be a slowly spinning black hole, on a timescale equal to the instability timescale. Thus the existence of rapidly spinning black holes in our universe constrains the instability timescale, which in turn constrains the mass of the field. Fig.~\ref{fig:bound2}, taken from Ref.~\cite{Pani:2012bp,Pani:2012vp}, shows how the observation of some black holes can constrain the mass of the photon to unprecedented levels. For further details, I refer the reader to Refs.~\cite{Pani:2012bp,Pani:2012vp,Witek:2012tr}.

There are two potential shortcomings of this approach. The first is that it assumes that the growth rate of the scalar field
is independent of nonlinear effects and that it grows until rotation is efficiently extracted out of the rotating black hole.
Once backreaction is taken into account, gravitational-wave emission (the flux of which scales with the square of the mass
of the boson cloud condensed outside the black hole, whereas the superradiant extraction scales linearly) can saturate the boson field-growth.
Effectively, this means that the scalar field might stop growing well before the superradiant condition $\mu\sim \omega \sim \Omega$
is satisfied. In other words, rotational energy extraction might be much smaller than that anticipated by
the superradiant instability. A thorough understanding of this effect requires fully nonlinear simulations, the first steps of which are currently underway \cite{Yoshino:2012kn,Okawa:2013}. 

The second shortcoming concerns the effects of matter around the black hole. It is reasonable to expect that matter, specially in thin accretion disks, cannot have a large effect in this instability as it is due to an asymptotic mass term and not tightly connected to equatorial properties.
However, an in-depth study is still lacking.
\subsection{Floating orbits}
Superradiance can also play an active role in the orbital evolution of binaries with 
a central black hole. To understand how an orbiting body around a black hole evolves,
one usually computes the energy budget, which is written as
\be
\dot{E}_{\rm binding}+\dot{E}_{\rm horizon}+\dot{E}_{\rm infinity}=0\,,
\ee
where $\dot{E}_{\rm binding},\,\dot{E}_{\rm horizon},\,\dot{E}_{\rm infinity}$ is the change in binding energy, flux at the horizon and at infinity,
respectively. Because the binding energy is related uniquely to the geodesic parameters, one is able to infer how the orbiting body evolves from one geodesic to another. Press and Teukolsky first suggested that it {\it might} be possible that the particle ``floats'' on a fixed circular geodesic 
for a very long time, provided that the flux at the horizon is negative and equal to that at infinity \cite{Press:1972zz}. These floating orbits are basically tapping the rotational energy of the black hole, in much the same way our moon is tapping Earth's rotational energy \cite{Cardoso:2012zn,Brito:2012gw}. Such orbits are not possible in pure GR \cite{Kapadia:2013kf}, but become possible as soon as 
massive scalar fields (or other bosonic fields) are included \cite{Cardoso:2011xi,Yunes:2011aa}. The boson mass introduces another scale in the problem and generically excitations with the same energy as the boson become possible: the instabilities are the result of exciting these quasi-particles close to the ergoregion, and the floating orbits as well. Generically, even without rotation long-lived states become possible.
Ultimately, both these resonances and floating orbits are possible due to the trapping of the scalar field: the mass term acts has an asymptotic wall at infinity, preventing radiation to escape.

It is important at this stage to stress that
minimally coupled bosons are interesting per se, but are not by far the most natural extension of GR. In this sense, both the superradiant instability
and floating orbits would be the analogue in gravitational physics of the resonant-state search that we routinely perform at particle accelerators.

Finally, the development of the energy extraction via floating orbits (by which I mean how the orbit evolvs on secular timescales) 
is not known and most likely requires self-force calculations.
\subsection{Generalized scalar-tensor theories and superradiance}
The inclusion of generic couplings of scalars to curvature lead to a whole new phenomenology. 
In particular, spontaneous scalarization and stronger superradiant effects are possible.
It was shown recently that the presence of matter in generic scalar-tensor theories leads to
the following equation for the scalars ~\cite{Cardoso:2013opa},
\be
\left[\square-\mu_s^2(x_i)\right]\varphi=0\,,\label{mass_dependent}
\ee
i.e., couplings of scalar fields to matter are equivalent to an effective position-dependent mass for the scalar field. 
It was also shown in Ref.~\cite{Cardoso:2013opa} that generic matter distributions may lead to strong superradiant instabilities in these theories.
Let me, for definiteness, consider the profile
\be
\mu^2_{0}= \frac{2\beta}{a^2+2r^2+a^2\cos2\theta}\Theta[r - r_0]\frac{(r-r_0)}{r^3}\,,\label{effective}
\ee
where $\Theta$ is the Heaviside function and I consider scattering off a Kerr BH with angular momentum $J=a/M$. This mass term separates the Klein-Gordon equation with the ansatz $\varphi=\Psi(r)S(\theta)e^{-i\omega t+im\phi}$. The angular eigenfunctions are the spin-weight-zero spheroidal wavefunctions \cite{Goldberg:1967sp,Berti:2005gp}, the eigenvalue is close to $l(l+1)$ with $l$ being an ``angular quantum number''.
\begin{table}[hbt]
\centering \caption{The gain coefficient for scattering of scalar waves in a matter profile ${\cal G}=\beta\Theta[r - r_0](r-r_0)/r^3$.} 
\vskip 12pt
\begin{tabular}{@{}cccccc@{}}
\hline \hline
&\multicolumn{5}{c}{${\rm Flux}_{\rm out}/{\rm Flux}_{\rm in}-1(\%)$}\\ \hline
$r_0$              &$\beta=500$  &$\beta=1000$   &$\beta=2000$            &$\beta=4000$        & $\beta=8000$\\
\hline \hline
%
%
$5.7$              &0.441            &0.604          &1.332                  &9.216                  &5.985$\times 10^{4}$\\
$6.0$              &0.415        &0.539          &1.059                   &5.589                 &513.2  \\
$10$               &0.369        &0.372          &0.380                   &0.399                 &0.825  \\
\hline \hline
\end{tabular}
\label{tab:amp}
\end{table}
The superradiant amplification factors are shown in Table~\ref{tab:amp} for selected values of $\beta$ and $r_0$.
The table shows the relative difference in ingoing and outgoing scalar flux at infinity,
$100\times ({\rm Flux}_{\rm out}-{\rm Flux}_{\rm in})/{\rm Flux}_{\rm in}$, which tells us how efficient superradiance is.
For small $\beta$ one recovers the standard minimally coupled results, with a maximum amplification of roughly $0.4\%$.
However, for certain values of $r_0,\beta$, the amplification factor can increase by several orders of magnitude,
making it a potentially observable effect.

\subsection{Ergoregion instability}
To end this overview on superradiant effects, I would like to mention a very important result by Friedmann \cite{Friedman:1978} which states that
spacetimes without horizons but with ergoregions are unstable. The physical explanation for this instability is that
a small negative-energy fluctuation in the ergoregion propagates outwards to infinity where it becomes positive energy.
By energy conservation, a larger negative-energy fluctuation must then develop in the ergo-region and so on and so forth.
The presence of an horizon can alleviate this by swallowing these negative energy states, which is why Kerr black holes 
are generically stable (against massless fields).

The ergo-region instability is also one efficient killer of black hole mimickers, i.e., of objects which are as compact and massive as black holes.
For those who are skeptical of the existence of black holes, mimickers are an interesting alternative. Well, the ergo-region instability
effectively rules out most of these alternatives as soon as they spin -- and most compact massive objects out there are spinning fast \cite{Cardoso:2007az,Cardoso:2008kj,Pani:2010jz}!

The connections between the ergo-region instability of certain string-theory inspired geometries \cite{Cardoso:2005gj} and Hawking radiation
were recently established by Mathur and co-workers \cite{Chowdhury:2007jx,Mathur:2012zp}.

\newpage
\section{Black hole collisions}
%
\begin{figure}[ht]
\begin{center}
\epsfig{file=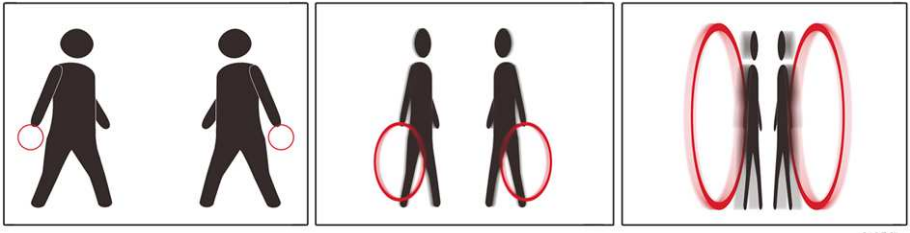,width=12cm,angle=0}
\caption{Black hole production from high energy collisions and the hoop conjecture.
Two objects and their respective hoops (of radii $R_{\rm hoop}=2GM\gamma/c^2$,
where $M$ is their rest-mass and $\gamma$ the Lorentz factor in the center-of-mass (CM) frame
of the collision) are shown for different velocities.
For large enough CM energy the objects fit inside the hoop and black hole production is possible.
\label{fig:hoop}}
\end{center}
\end{figure}
Black hole collisions were traditionally studied with the aim of understanding gravitational-wave emission from astrophysical sources
and their subsequent detection in Earth-based detectors. The generalized interest in trans-Planckian collisions and black hole formation
at very high energies has renewed the interest in this topic. 
The idea is very simple and has two logical steps: (i) collisions of particles at extremely high energies 
give rise to a black hole as final state; (ii) at very high energies, collisions of particles are well described by collisions of black holes.
Thus, the outcome of these two logical steps is that trans-Planckian collisions of particles
are well described by black hole collisions, as long as the outcome is a single black hole.
In practice, this amounts to reducing the problem to the study of a 1-parameter family of initial data (impact parameter), a prodigious simplification.

How well-justified are the underlying assumptions? Assumption (i) is supported by the Hoop Conjecture \cite{Thorne:1972}
which roughly states that two objects of mass $2M_{\rm tot}$ form a black hole if a hoop of radius $4GM_{\rm tot}/c^2$
can be made to pass around the objects in all possible directions. Because all forms of energy gravitate, the total mass $M_{\rm tot}$
should include kinetic energy, $M_{\rm tot}=M\gamma$, where $M$ is the rest-mass energy and $\gamma$ a Lorentz factor.
The evolution of the hoop radius as the boost increases is depicted in Fig.~\ref{fig:hoop}: because the typical size of the object does not increase
with boost, at some critical Lorentz factor the object {\it will} fit inside the hoop. In conclusion, high-energy collisions of particles should 
produce black holes for very large CM energies. The hoop conjecture is by no means proven, but recent results seem to support it.
Refs.~\cite{Choptuik:2009ww} and \cite{East:2012mb,Rezzolla:2012nr} studied the collision of two objects (boson and fluid stars respectively) as the CM energy increases, and found that the critical Lorentz boost predicted by the hoop conjecture {\it overestimates} the critical boost seen in their simulations.

\begin{figure}[t]
\begin{center}
\begin{tabular}{cc}
\epsfig{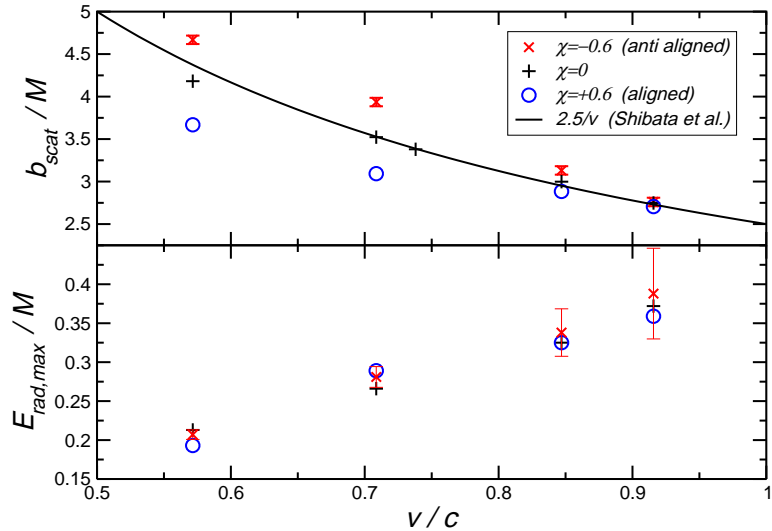} &
\end{tabular}
\caption{Critical scattering threshold (upper panel) and maximum
  radiated energy (lower panel) as a function of $\gamma$. Blue
  circles and red crosses refer to the aligned and antialigned case,
  respectively. Black ``plus'' symbols represent the thresholds for
  the four nonspinning configurations, complemented (in the upper panel) by results from
  \cite{Sperhake:2009jz} for $\gamma=1.520$. For clarity, we only plot
  error bars for the antialigned-spin sequence. Taken from Ref.~\cite{Sperhake:2012me}.
  \label{fig:d4bscat}}
\end{center}
\end{figure}

The assumption (ii) is based on the fact that once an horizon forms no information gets out. Thus, all the information about the multipolar structure
of the object is forever hidden behind the horizon. {\it Matter does not matter} is the catchphrase used to describe this.
The correctness of this assumption is also borne out of recent investigations \cite{East:2012mb,Sperhake:2012me}. East and Pretorius \cite{East:2012mb} collided two fluid stars at very high energies (such that a black hole is produced), and have convincing evidence that the gravitational waveforms and spectra are very similar to that from the collision of two black holes \cite{Sperhake:2008ga,Sperhake:2009jz}.
Perhaps more impressive is the recent work by Sperhake et al \cite{Sperhake:2012me}. To test how the multipolar structure is irrelevant at large energies, they have taken spinning black holes (either aligned or anti-aligned) and collided them at progressively higher energies. By measuring the critical impact parameter for scattering, they can see the effect of spin and CM energy. The results are summarized in Fig.~\ref{fig:d4bscat}.
This figure is very clear: at large CM energies, details about spin (whether it is aligned, anti-aligned or zero) become irrelevant,
the total collision energy is the controlling factor.

To summarize, all available data suggests that black hole formation from very high energy particle collisions
can be modelled using non-spinning black hole-black hole collisions. 
\subsection{On the unreasonable effectiveness of approximation tools}
%
\begin{figure*}[htb]
\begin{center}
\begin{tabular}{c}
\includegraphics[scale=0.4,clip=true]{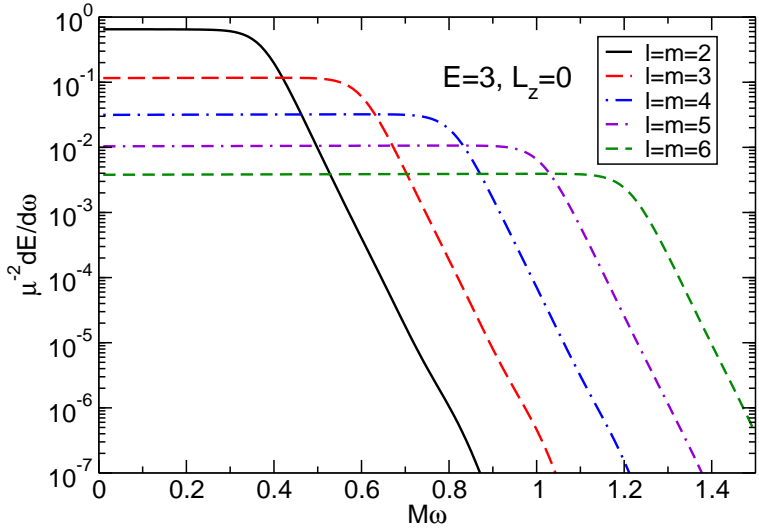} 
\end{tabular}
\end{center}
\caption{\label{E1L0} Spectra for $l=m=2,\dots,6$ for kinetic energy-dominated infalls ($\gamma=3$). 
Taken from Ref.~\cite{Cardoso:2002ay,Berti:2010ce}.}
\end{figure*}
In this subsection, I focus entirely on head-on collisions for the sake of concreteness. 
There are several approaches to understand black hole collisions.
One possibility consists in studying first an extreme mass ratio binary
in which a large black hole of mass $M$ collides with small black hole of mass $\mu\ll M$.
This process can be studied perturbatively in $\mu/M$ \cite{Davis:1971gg,Cardoso:2002ay,Berti:2010ce,Sperhake:2011ik,East:2013iwa},
and the total integrated radiated energy in gravitational waves is found to be
\be
\frac{E}{M}=0.0104\frac{\mu^2}{M^2} \,\,(\gamma=1)\qquad   {\rm and}\,\,\,\, \frac{E}{M}=0.26\frac{\gamma^2\mu^2}{M^2} \,\,(\gamma\to\infty)\,.
\ee
For rest mass-dominated collisions $\gamma\sim 1$, it seems sensible to extrapolate to arbitrary mass ratios by promoting $\mu$
to the system's reduced mass and $M\to M_{\rm tot}$ to the total mass. This yields $E/M_{\rm tot}\sim 0.00065$ 
in very good agreement with numerical results \cite{Sperhake:2011ik,Witek:2010xi}. 
For kinetic energy-dominated collisions one possible prescription is to let $\mu,M \to M_{\rm tot}/2$, yielding an efficiency of roughly $E/M_{\rm tot}\sim0.13$ for gravitational 
wave generation, in remarkable agreement with latest numerical results \cite{Sperhake:2008ga,East:2012mb} which quote $E/M_{\rm tot}\sim 0.14$.

Adding the first $10$ multipoles (of a high-energy point particle falling into a massive black hole), I find that the total flux peaks when the particle crosses the light-ring, and that
\be
\dot{E}_{\rm peak} \sim 0.014\frac{\mu^2}{M^2}\,.
\ee
Extrapolating to the equal mass case with the prescription $\mu\to M\to M_{\rm tot}/2$, I estimate that the peak luminosity in the head-on collision
of two equal mass black holes reaches $\dot{E}_{\rm peak}\sim 0.015$. This estimate is in excellent agreement with actual nonlinear numerical calculations \cite{Sperhake:2008ga,Sperhake:2009jz}. The peak flux from such collisions surpasses the largest luminosity 
from any known process in nature. The interesting thing about such numbers is that they allow to test the conjecture
that there is a limit of $\sim c^5/G$ for the maximum possible luminosity \cite{thorne}
\footnote{
This bound has an interesting story. Kip Thorne, and others after him, attribute the conjecture to Freeman Dyson; I learned recently that Freeman Dyson denies he ever made such conjecture \cite{gibbons}, and instead attributes such notion
to his 1962 paper \cite{Dyson}; in that work he shows that the power emitted by a binary
approaches the upper limit (and mass-independent number)$128c^5/5G$ as the velocity of the binary members $v\to c$.  
}. As far as I'm aware no process has ever been found to yield larger luminosities than this bound, but high energy black hole collisions
come pretty close.

\begin{figure}[htb]
\begin{center}
\includegraphics[scale=0.33,angle=270,clip=true]{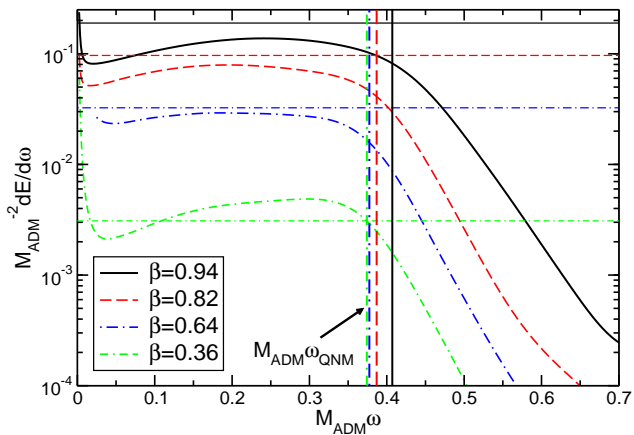}
\end{center}
\caption{\label{fig:headonspec} Energy spectrum for the dominant (quadrupolar,
  i.e. $l=2$) component of the gravitational radiation computed from NR
  simulations of the head-on collision of two equal-mass black holes (from
  \cite{Sperhake:2008ga}). The collision speed in the center-of-mass frame,
  $\beta=v/c$, is indicated in the legend. The energy spectrum is roughly flat
  (independent of frequency) up to the quasinormal mode (QNM) frequencies
  (marked by vertical lines), after which it decays exponentially. All
  quantities are normalized to the Arnowitt-Deser-Misner (ADM) mass of the
  system $M_{\rm ADM}$.}
\end{figure}
An alternative method to understand high-energy black hole collisions is known as the Zero Frequency Limit
calculation pioneered by Weinberg and Smarr \cite{Weinberg:1964ew,Weinberg:1965nx,Adler:1975dj,Smarr:1977fy,Berti:2010ce}. 
This approach is well-known also in electrodynamics \cite{jackson}. The ZFL models the collision by taking two constant-velocity particles
which instantaneously collide at the instant $t=0$.
The lack of any lengthscale in the problem and the existence of an infinite acceleration produces a formally divergent 
total radiated energy, a problem which can be remedied by introducing a physical cutoff. This physical cutoff is the lowest quasinormal frequency of the final black hole.

The ZFL predicts a flat spectrum, which {\it coincides} (within numerical precision) to the linearized calculation
of a small particle falling into a massive black hole. The latter are shown in Fig.~\ref{E1L0}, where it is clear that the black hole does introduce
a lengthscale and a cutoff, very accurately described by the lowest quasinormal frequency. To summarize, the ZFL predicts the correct 
behavior for the spectra. How well does it describe nonlinear simulations? A typical example is shown in Fig.~\ref{fig:headonspec}, where we compare
the ZFL prediction with the spectra from the collision of two equal-mass black holes colliding at high velocities. Even in this highly nonlinear process,
perturbative schemes capture very well the fine details of the waveforms.

The unique aspect of these extrapolations is not that they get the correct number to a few percent (that is simply astonishing), it is rather the fact that such naive
extrapolations work in a wide range of kinetic energies up to factors of order two, whichever way the extrapolation is done. This ``unreasonable effectiveness'' of approximation theories (to borrow a line from Cliff Will in the context of Post-Newtonian expansions, which in turn is an adaptation from one of Wigner's papers \cite{Cliff:Chandra,Wigner}) is seen also in the fine details of the waveforms and in other approximation techniques. It seems non-linearities are either red-shifted away or eaten by the newly-formed black hole. Such washing away of non-linearities have been observed in other spacetimes \cite{Witek:2010xi,Witek:2010az,Bantilan:2012vu} and certainly deserves further investigation.
\subsection{Anatomy of close fly-by's}
%
\begin{figure*}[tb]
\begin{center}
\begin{tabular}{c}
\epsfig{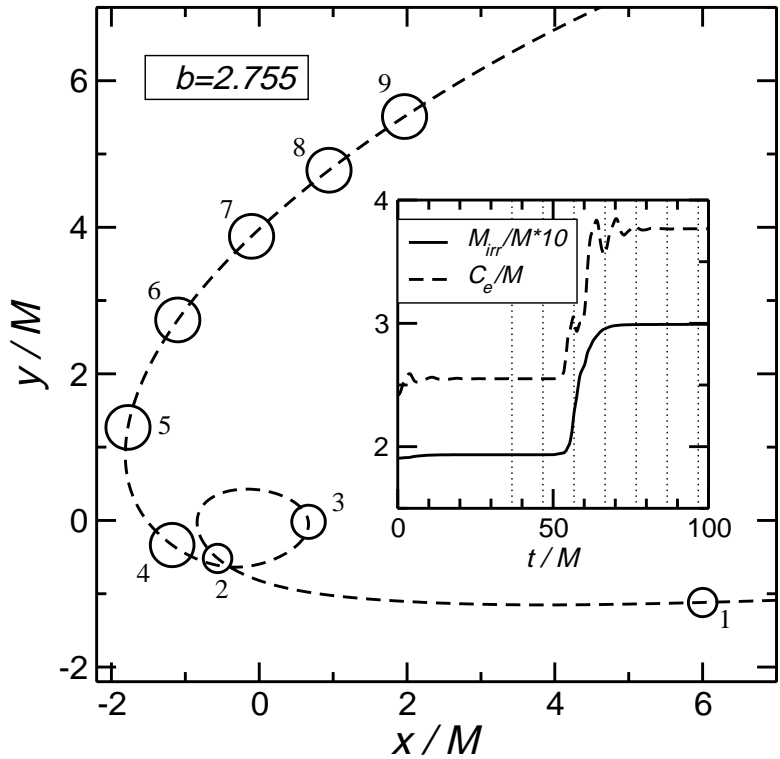} 
\end{tabular}
\caption{
  Trajectory of
  one black hole from the simulation with $b/M=2.755$. Inset: time
  evolution of the irreducible mass $M_{\rm irr}$ and of the
  circumferential radius $C_{\rm e}$ of each hole. The circles
  represent the black hole location at intervals $\Delta t=10~M$
  (corresponding to vertical lines in the inset) and have a radius equal
  to $M_{\rm irr}$. Taken from Ref.~\cite{Sperhake:2012me}.}
\label{fig:triple}
\end{center}
\end{figure*}
Let me finalize this discussion with a unique feature of non head-on collisions.
When two black holes are shot at one another at close to the speed of light, they can orbit more than once 
before either merging or scattering. These orbits are reminiscent of unstable geodesics and some of them display a ``zoom-whirl''
type of behavior \cite{Pretorius:2007jn,Sperhake:2009jz,Shibata:2008rq}. Fig.~\ref{fig:triple} summarizes what happens when the end-result is a scatter. The two black holes were initially spinning with $a/M=0.6$ ~\cite{Sperhake:2012me}. 
A large amount of radiation is generated during the process, which subsequently gets (partly) absorbed by the scattered black holes;
on absorbing these large amounts of energy, the scattered black holes grow in size and mass.
We estimate that for very large CM energies, roughly half the CM energy is absorbed by the black holes, implying that the fine-tuned (in impact parameter) collision of two black holes can never dissipate $100\%$ of the CM energy as gravitational waves. Again, perturbative techniques point in the same direction
\cite{Sperhake:2012me,Gundlach:2012aj}.
As a final remark, zoom-whirl behavior seems to be intrinsic to, or at least more colourful in, four-dimensional spacetimes \cite{Okawa:2011fv}.
	
\section{Conclusions}
The next decades hold the promise to uncover the strong field region 
close to black holes and neutron stars. Our efficiency to explain or interpret future observations
depends strongly on our ability to understand these objects; there is reason to be optimistic, as the last years witnessed a real breakthrough
in nonlinear time evolutions and powerful approximation techniques. 
In parallel, developments on the theory side show that black holes can be used to understand high energy physics. The no-hair theorem
guarantees that black holes are the simplest objects one can collide to investigate the outcome of very high-energy collisions;
curiously, astrophysical black holes can also be used to probe particle physics;
the possibility of using black holes as particle detectors, either by resonances in floating orbits or in spacetime instabilities
is also intriguing...It is now up to experiments to show us the way.

\section{Acknowledgements}
This work is the result of direct interactions with many colleagues throughout the last years. For discussions, correspondence and collaboration
I am specially indebted to Enrico Barausse, Emanuele Berti, Richard Brito, \'Oscar Dias, Sam Dolan, William East,
Leonardo Gualtieri, Carlos Herdeiro, Luis Lehner, Jos\'e Lemos, Madalena Lemos, Akihiro Ishibashi, Hideo Kodama, Hirotada Okawa, Miguel Marques, Paolo Pani, Frans Pretorius, Jorge Pullin, Thomas Sotiriou, Ana Sousa, Ulrich Sperhake, Jo\~ao Rosa, Helvi Witek, Shijun Yoshida, Hirotaka Yoshino, Nicolas Yunes and Miguel Zilh\~ao. All the figures were designed by Ana Sousa.

I also thank all the participants of the ``Extra-Dimension Probe by Cosmophysics'' workshop in KEK, the YITP-T-11-08 workshop on ``Recent advances in numerical and analytical methods for black hole dynamics,'' the Perimeter Institute workshop ``Exploring AdS-CFT Dualities in Dynamical Settings,''
the ``Spanish Relativity Meeting 2012,'' and the Helsinki DECI Minisymposium Workshop on State of the Art in Scientific and Parallel Computing. 
%
Finally, I acknowledge partial financial
support provided under the European Union's FP7 ERC Starting Grant ``The dynamics of black holes:
testing the limits of Einstein's theory'' grant agreement no.
DyBHo--256667 and the NRHEP 295189 FP7-PEOPLE-2011-IRSES Grant.
Research at Perimeter Institute is supported by the Government of Canada 
through Industry Canada and by the Province of Ontario through the Ministry
of Economic Development and Innovation.


\end{document}